\documentclass[epj]{svjour}
\usepackage[utf8x]{ inputenc }
\usepackage[T1]{ fontenc }
\usepackage{ ae, aecompl }

\usepackage{ amsmath, amssymb, amsfonts, amsbsy, bbm, bm, ulem }
\usepackage{color}
\usepackage{graphicx}
\usepackage{grffile}
\usepackage{nath}
\def\abs#1{\left|#1\right|}

\newcommand{\ave}[1]{\langle#1\rangle}

\renewcommand{\operatorname}[1]{{`#1}}
\renewcommand{\vec}[1]{\mathbf{#1}}

\def\nexspherio{NexSPheRIO}

\def\identity{{\rm 1\hspace{-2.5pt}l}}

\def\ave#1{\langle#1\rangle}

\def\std#1{{`std}\{#1\}}
\def\lim{{`lim}}

\def\det{{`det}}

\def\m2{\{2\}}

\newcommand{\myfigure}[1]{%
\begin{figure}%
\begin{center}%
#1%
\end{center}%
\end{figure}%
}

\begin{document}

\title{
Coarse graining scale and effectiveness of hydrodynamic modeling
}

\author{
Ph. Mota\inst{1,2}, T. Kodama\inst{1,2,3}, R. Derradi de Souza\inst{4} \and J. Takahashi\inst{4}
}
\institute{
Instituto de Física, Universidade Federal do Rio de Janeiro,  Av. Athos da Silveira Ramos 149, 21941-972 Rio de Janeiro, Brazil
\and
Frankfurt Institute for Advanced Studies FIAS, Goethe-Universität Frankfurt, Ruth-Moufang-Str. 1, 60438 Frankfurt am Main, Germany
\and
ExtreMe Matter Institute EMMI and Research Division, GSI Helmholtzzentrum für Schwerionenforschung, Planckstraße 1, 64291
Darmstadt, Germany
\and
Instituto de Física Gleb Wataghin, Universidade Estadual de Campinas, Rua Sérgio Buarque de Holanda 777, 13083-859 São Paulo, Brazil
}

\abstract{
Some basic questions about the hydrodynamical approach to relativistic heavy ion collisions are discussed aiming to clarify how far we can go with such an approach to extract useful information on the properties and dynamics of the QCD matter created.
We emphasize the importance of the coarse-graining scale required for the hydrodynamic modeling which determines the space-time resolution and the associated limitations of collective flow observables.
We show that certain kinds of observables can indicate the degree of inhomogeneity of the initial condition under less stringent condition than the local thermal equilibrium subjected to the coarse-graining scale compatible to the scenario.
}

\maketitle
\section{Introduction}

Hydrodynamic approach has shown to be very successful to describe the global
and collective features of the process in relativistic heavy ion
collisions. In particular, the behavior of elliptic flow parameter
$v_{2}$ as function of centrality and transverse momenta data is well
reproduced by hydrodynamic models \cite{PhysRevLett.86.402,Adams2005102,Adcox2005184,PhysRevC.72.051901,PhysRevC.80.024909}. It should be mentioned that other
approaches based on binary collisions of constituent particles like parton
cascade in general yield smaller collective flow values than the observed values.
This success of hydro approach leads to the expectation that, from a detailed
hydrodynamic analysis of experimental data, we may be able to determine the
thermal properties of QCD matter such as the equation of state (EoS) and
transport coefficients, which can be then compared to those obtained from the
lattice QCD (\textit{l}QCD) calculations. In fact, many works have been and
are being done in this direction. Of course, these are valid and important
efforts that must be done to verify how far we can go with a given working
hypothesis, that is, the validity of hydrodynamics.

On the other hand, these successes brought us several new interesting
questions and mysteries such as very early thermalization.
The most crucial one is that why at all the hydrodynamic approach works so well for such a violent and almost microscopic collisional process.
It is commonly believed that the basic hypothesis of hydrodynamics
is the validity of {\it local thermodynamical equilibrium}.
That is, in every space-time points, all the thermostatic state variables
are well-defined and associate thermodynamic equations are valid locally.
If this is true, and the hydrodynamic analysis is giving a unique scenario, then we are led to
conclude that the thermalization time and correlation length are really
extremely small for the QCD matter produced in the nuclear collisions.
This is an important consequence for further understanding of the QCD
matter at extreme conditions.

Note that in the above statement, there are two important `{\emph{if\/}}'s,
that is, `if really the hydrodynamics necessarily implies the local thermal
equilibrium' and `if the present analysis's can be considered as a unique
solution for the observed collective behavior of the relativistic heavy ion
collisions'. In this paper, we would like to address these `if's, and discuss
what will be the experimental measure to estimate the validity of the
hydrodynamic statements. In particular, it is fundamental to discover
observables which are sensitive to the local dynamics of the system. We
organize the present paper as follows. In Sec. II, for the sake of later
bookkeeping, we make a brief review of the structure of the relativistic
hydrodynamics keeping the above two if's in mind, in particular, from the view
point of the variational principle. We then discuss how the concept of
coarse-graining can be introduced in the derivation of the hydrodynamic
equations. It will be shown that the ideal hydrodynamic description can be
considered as an effective model in terms of course-grained field variables.
In Sec. III, we discuss the compatibility between the coarse-graining scale
and observables. We argue that most of collective flow signals, if averaged
over events, are insensitive to the detailed space-time structure of the
dynamics, as far as the coarse-graining scale is compatible with these
observables. The genuine local hydrodynamic signal with local thermodynamic
equilibrium requires high resolution on the space-time evolution, and such
an information can only be meaningful for each collisional event and not for
the averaged observables. In this work, using a controlled set of initial
conditions with different granularity, we look for observables which reflect
the event-by-event (EbE) dynamical information and consequently are sensitive
to the  initial state inhomogeneity.  Section IV is devoted to the discussion
of our results and perspectives.

\section{Relativistic Hydrodynamics}

In this section, we review the derivation of relativistic hydrodynamics
(mainly for the ideal case) and discuss the meaning of local thermal
equilibrium in the context of coarse-graining procedure.

First, let us consider a (classical) matter carrying some conserved quantity
$N$, to which four-current density $n^{\mu}(x)$ satisfies the continuity
equation,
\begin{equation}
\partial_{\mu}n^{\mu}(x)=0.\label{continuity-n}%
\end{equation}

In this note, to avoid unnecessary complexity for those who are not used to
the \textquotedblleft general relativistic\textquotedblright\ notations, we
consider only the case of Minkowsky coordinate, $x=\{x^{0},\vec{x}\}$ with the metric
$g_{\mu\nu}=\operatorname{diag}\{1,-1,-1,-1\}$.
The energy-momentum tensor $T^{\mu\nu
}(x)$ of the system also conserves,
\begin{equation}
\partial_{\mu}T^{\mu\nu}(x)=0.\hspace{1em}\label{divTmunu}%
\end{equation}
In the case that $T^{\mu\nu}(x)$ is a $4\times4$ symmetric matrix, {\it e.g.} in the absence of electromagnetic fields or spin variables, we can diagonalize at any point $x$.
From the physical requirement, we know that there is only one time-like eigenvector $u^{\mu}$, with positive eigenvalue
$\varepsilon$ \cite{LandauLifshitz},
\begin{equation}
T^{\mu\nu}u_{\nu}=\varepsilon u^{\mu}.\label{Eigen}%
\end{equation}
We can identify this above time-like eigenvector, with proper normalization,
\begin{equation}
u_{\mu}u^{\mu}=1,\label{u-normal}%
\end{equation}
as the four-velocity for the Lorentz transformation from the observable system
to the local rest frame of energy flow.\ Consequently, the eigenvalue
$\varepsilon$ is the proper energy density, that is, the energy density
measured in the rest frame of the energy flow at the space-time position $x$.
The rest frame of the energy flow is called Landau-Lifshitz frame (LL). In LL frame, the
energy-momentum tensor becomes:
\begin{equation}
T^{\mu\nu}\rightarrow[
\begin{matrix}
\varepsilon & \boldsymbol{0}\\
\boldsymbol{0} & \mathbb{T}
\end{matrix}
]  ,\label{TmunuLocal}%
\end{equation}
where $\mathbb{T}$ is the $3\times3$ stress tensor.

In general, the direction of the energy flow vector does not necessarily
coincide with that of the matter current, that is, not always $n^{\mu}\propto
u^{\mu}$. In such cases, we write
\begin{equation}
n^{\mu}=nu^{\mu}+q^{\mu},\label{nmuu-q}%
\end{equation}
where we choose $q^\mu n_\mu=0$ so that $n$ represents the density of
the conserved quantity $N$ measured in LL frame and $q^{\mu}$ is the
diffusion current with respect to this frame. For the sake of later
convenience, we introduce the following projection operators,
\begin{equation}
\Delta_{\parallel}^{\mu\nu}=u^{\mu}u^{\nu},\label{Deltapara}%
\end{equation}
and
\begin{equation}
\Delta_{\perp}^{\mu\nu}=g^{\mu\nu}-u^{\mu}u^{\nu},\label{Delta_ortho}%
\end{equation}
which can be used to decompose vectors or tensors systematically into the
parallel and perpendicular components to the four - velocity $u^{\mu}$. For
example, the above Eq.(\ref{nmuu-q}) is expressed formally as
\begin{equation}
n^{\mu}=(\Delta_{\parallel}^{\mu\nu}+\Delta_{\perp}^{\mu\nu})n_{\nu}=(n^{\nu
}u_{\nu})u^{\mu}+\Delta_{\perp}^{\mu\nu}n_{\nu},\label{decom_n}%
\end{equation}
and identify
\begin{equation}
n=u_{\mu}n^{\mu},q^{\mu}=\Delta_{\perp}^{\mu\nu}n_{\nu}.\label{nandq}%
\end{equation}
These projection operators can be used also to decompose the energy-momentum tensor. Due to the symmetry of $T^{\mu\nu}$, in LL frame, we can write as
\begin{equation}
T^{\mu\nu}=\varepsilon\Delta_{\parallel}^{\mu\nu}+\Delta_{\perp}^{\mu\alpha
}\Delta_{\perp}^{\nu\beta}T_{\alpha\beta}\label{Tmunu-decomp}%
\end{equation}
where the last term corresponds essentially to the $3\times3$ stress tensor
$\mathbb{T}$ in the local rest frame. We can further decompose the tensor
$\mathbb{T}$ into isotropic irreducible parts of the local rotational symmetry as
\begin{equation}
\mathbb{T}=\left.\frac{1}{3}\operatorname{tr}(\mathbb{T})\identity\right.
+[\mathbbm{T}-\frac{1}%
{3}\operatorname{tr}(\mathbbm{T})\mathbbm{1}]\label{T3Ddecomp}%
\end{equation}
where $\mathbbm{1}$ is the $3\times3$ identity matrix. This decomposition results
in the original energy-momentum tensor as
\begin{equation}
T^{\mu\nu}=\varepsilon\Delta_{\parallel}^{\mu\nu}-P\Delta_{\perp}^{\mu\nu}%
+\Pi^{\mu\nu},\label{HydroTmunu}%
\end{equation}
where
\begin{equation}
P=-\frac{1}{3}\Delta_{\perp}^{\alpha\beta}T_{\alpha\beta}%
,\label{DynmicalPress}.%
\end{equation}
Note that at this point $P$ is the dynamical pressure which is the sum of the hydrostatic pressure and bulk viscosity.
And
\begin{equation}
\Pi^{\mu\nu}=\Delta_{\perp}^{\mu\alpha}\Delta_{\perp}^{\nu\beta}T_{\alpha
\beta}-\frac{1}{3}\Delta_{\perp}^{\alpha\beta}T_{\alpha\beta}\Delta_{\perp
}^{\mu\nu},\label{Pimunu}%
\end{equation}
comes from the non-diagonal elements of the local stress tensor. $\Pi^{\mu\nu
}$ is a symmetric tensor, satisfying
\begin{equation}
\Pi_{\mu}^{\mu}=0,u_{\mu}\Pi^{\mu\nu}=u_{\nu}\Pi^{\mu\nu}=0.\label{Pi_ortho}%
\end{equation}
At this stage, the original $4\times4$ symmetric tensor $T^{\mu\nu}$ (10
parameters) is expressed in terms of $u^{\mu},\varepsilon,P$ and $\Pi^{\mu\nu
}$ (10 = 3+1+1+5). Together with the conserved current $n^{\mu}$, we have to
know the time evolution of these 14 variables when their initial values are
given. When the time-evolution equations form a closed system within these 14
variables, then we have defined the system of partial differential equations  of (dissipative) relativistic hydrodynamics.

\subsection{Ideal Fluid}

In some special physical situations, the total number of variables reduces
drastically. Suppose that the 3 eigenvalues of the local stress tensor are
degenerated. Then $\mathbb{T}$ becomes isotropic and consequently the tensor
$\Pi^{\mu\nu}$ vanishes. Furthermore, if there is no diffusion of the
conserved current in LL frame, then $q^{\mu}$ also vanishes. In such a
situation, the total number of variables reduces to 6. Since the conservation
laws, Eqs.(1) and (2) furnish 5 equations among them, we need only one
equation to close the system. Usually we introduce the so-called Equation of
State (EoS) which establishes a functional relation among the local
quantities, $\varepsilon,P,n$, as
\begin{equation}
P=P(  \varepsilon,n)  ,\label{P(e,n)}%
\end{equation}
which completes the system of ideal hydrodynamic evolution.
Note that in such a system, the dynamical pressure is identified as the hydrostatic pressure.
In other words, there is no bulk viscosity.
More explicitly,
for the set of 5 independent variables, $\varepsilon,n,\vec{u}$,
\ Eqs.(\ref{continuity-n}) and (\ref{divTmunu}) can be written in the form of
time evolution equations as
\begin{equation}
\gamma \frac{d}{dt}\varepsilon = - ( \varepsilon+P )\partial_\mu u^\mu,\label{dedt}%
\end{equation}%
\begin{equation}
\gamma \frac{d}{dt} n = -n \partial_\mu u^\mu,\label{dndt}%
\end{equation}
\begin{equation}
\gamma n \frac{d}{dt}( \frac{\varepsilon+P}{n} u^i )  = -\partial^i P,\label{EulerRel}%
\end{equation}
where $\frac{d}{dt}\varepsilon = \partial_{0}\varepsilon + v^i \partial_i \varepsilon$.
Roman letters were used to denote spatial only indexes ($i=1,2,3$).
This set of equations can be solved together with Eq. (\ref{P(e,n)})
for a given initial condition for $(\varepsilon,n,{v})$.
In the equations above, $\gamma=u^{0}$ is the
Lorentz factor, ${v}^i={u}^i/\gamma$ and $n^{\ast}=\gamma n$ are,
respectively, the (three) velocity and the density in the global observer's
system. From now on, we use the convention that the symbol $^{\ast}$ is used to
distinguish the value defined in the global observational system of the
corresponding quantity without $^{\ast}$ defined in the local rest frame.

\subsection{Variational Principle}

We see that the hydrodynamic equation for an ideal fluid can be
obtained from the continuity equations Eqs. (\ref{continuity-n}) and
(\ref{divTmunu}) with additional assumptions such as local isotropy of the
stress tensor, null diffusion of the conserved quantity in LL frame
and the existence of the EoS. All these additional conditions can be satisfied
if the thermodynamical equilibrium is satisfied locally. For this reason, it
is commonly believed that the success of hydrodynamic description in
relativistic heavy ion collisions indicates that local
thermodynamical equilibrium is attained in these processes.

In fact, the set of hydrodynamic equations formally constitutes a local
covariant classical field theory. If we consider that relativistic
hydrodynamics as in a covariant theory, then the local thermal equilibrium should
be attained in each space-time point. This is somewhat a physically
contradicting condition, because to attain the thermal equilibrium, we need a
large volume and time. Furthermore, even if we admit that the local thermal
equilibrium could be the true answer for the success of the hydrodynamic
approach in the relativistic heavy ion collisions, we don't know yet any
quantitative measure to determine the precision of the hydro predictions for
the properties of the matter in question. Here, to further clarify these
questions, it may be useful to derive the equations of motion for an ideal
fluid dynamics from a different point of view.

Consider a small volume element $\Delta V^{\ast}$ of the fluid at the
position $\vec{x}$, whose velocity is $\vec{v}$. The relativistic Lagrangian
of a particle with the rest mass $m$ is given by
\begin{equation}
L=-m/\gamma\label{L_particle}%
\end{equation}
where $\gamma$ is the Lorentz factor of the particle. The four-velocity of the
particle is given by
\begin{equation}
(  u^{\mu})  =(
\begin{array}
[c]{c}%
\gamma\\
\gamma\vec{v}%
\end{array}
)  ,\label{four-velocity}%
\end{equation}
as usual.

Considering the fluid element as a particle, the intrinsic volume is $\Delta
V=\Delta V^{\ast}/\gamma$ and the (average) energy density $\varepsilon$ in
its rest frame is related to the rest mass of this piece of fluid as
$m=\varepsilon\gamma\Delta V^{\ast}$. The Lagrangian of the fluid is then
written as
\begin{equation}
L=\underset{x}{-\sum}\varepsilon\Delta V^{\ast}\label{Lagrange_sum}%
\end{equation}
where the summation is taken over every fluid element.
In the limit of small volume element, we can write
\begin{equation}
L=-\int d^{3}x\ \varepsilon.\label{Continuum Lagrange}%
\end{equation}
We suppose that in each fluid element, the proper energy density $\varepsilon$
can be expressed as function of the proper density $n$ of the conserved
quantity as
\begin{equation}
\varepsilon=\varepsilon(  n)  .\label{e(n)}%
\end{equation}
By definition, the current, $n^{\mu}\equiv nu^{\mu}$ conserves,
\begin{equation}
\partial_{\mu}n^{\mu}=0.\label{cont_n}%
\end{equation}
Let us then apply the variational principle for the action,
\begin{equation}
I=-\int d^{4}x\ \varepsilon \label{action0}%
\end{equation}
together with the constraints, Eq.(\ref{cont_n}) and the normalization
condition of the four-velocity,
\begin{equation}
u_{\mu}u^{\mu}=1.\label{u_nor}%
\end{equation}
Introducing Lagrange multipliers for the constraints, we get
\begin{equation}
\delta\int d^{4}x\left[  \varepsilon+\lambda(  \partial_{\mu}n^{\mu
})  +\xi(  u_{\mu}u^{\mu}-1)  \right]
=0,\label{Variation_Euler}%
\end{equation}
for the variations of $(  n,u^{\mu},\lambda,\xi)$. It can be
shown \cite{JPhysG.25.1935} that the variational procedure leads exactly to the
conservation of the energy-momentum tensor,
\begin{equation}
\partial_{\mu}T^{\mu\nu}=0,\label{divTmunu_2}%
\end{equation}
where $T^{\mu\nu}=(  \varepsilon+P)  u^{\mu}u^{\nu}-Pg^{\mu\nu}$
with
\begin{equation}
P=n\frac{\partial\varepsilon}{\partial n}-\varepsilon.\label{P=dedn}%
\end{equation}
Thus, the variational principle with the action Eq. (\ref{action0}), with the
constraints, Eqs. (\ref{cont_n}) and (\ref{u_nor}) leads to the ideal
hydrodynamic equations, Eqs. (\ref{dedt}-\ref{EulerRel}). Here, we showed for
simplicity the case of only one conserved quantity, but it is straightforward
to generalize when there are more conserved quantities, say, $\left\{
n_{i},i=1,\ldots,r\right\}$. We find that everything is the same, except
for the definition of the \textquotedblleft pressure\textquotedblright, Eq.
(\ref{P=dedn}) to
\begin{equation}
P=\underset{i}{\sum}n_{i}\frac{\partial\varepsilon_{i}}{\partial
n}-\varepsilon.\label{P_geral}%
\end{equation}
In the above we showed that the set of ideal hydrodynamic equations can be
derived from a variational principle with the action Eq.(\ref{action0}).
There, the most fundamental condition is that the local energy density can be
expressed as function of other extensive conserved quantities,
\begin{equation}
\varepsilon=\varepsilon(  n_{1},\ldots,n_{r})  .\label{e_geral}%
\end{equation}

\subsection{Variation in Lagrangian Coordinates}

For the sake of later discussion, we will repeat the derivation of
hydrodynamic equations using the Lagrangian coordinates. Let us introduce the
space coordinate system $\left\{  \vec{X}\right\}  $ at the initial time
$t_{0}$. We can write the fluid density distribution at this instant as
\[
n_{0}=n_{0}(  \vec{X})  .
\]
The fluid element initially at the position $X$ moves according to the
equation of motion and occupies the other location, say $\vec{x}=\vec
{r}(  t;X)  $. Here, $\vec{r}(  t;X)  $ represents the
trajectory of the fluid element initially located at $\vec{X}$. Thus, the
fluid configuration at the instant $t$ in the observer's system is written as
\begin{equation}
n^{\ast}(  \vec{x},t)  =\int d^{3}\vec{X}\ n_{0}^{\ast}(
\vec{X})  \delta^{3}(  \vec{x}-\vec{r}(  t;\vec{X})
)  ,\label{dens}%
\end{equation}
which can be rewritten as
\begin{equation}
n^{\ast}(  \vec{x},t)  =\frac{1}{J}n_{0}^{\ast}(  \vec
{X})  ,\label{Jacob}%
\end{equation}
where
\begin{equation}
J=\det(  \frac{\partial\vec{r}}{\partial\vec{X}})  _{\vec{x}%
=\vec{r}}.\label{J}%
\end{equation}
The local proper density $n$ is then expressed as%
\begin{equation}
n=\frac{1}{\gamma}n^{\ast}=\frac{n_{0}^{\ast}}{\gamma\ J}.\label{proper_n}%
\end{equation}
When $n_{0}^{\ast}$ is a smooth function\footnote{See the later discussion
when $n_{0}^{\ast}$ is not a smooth function.}, by a suitable variable
transformation, we can consider $n_{0}^{\ast}$ as constant without losing generality \cite{1751-8121-45-25-255204}.

The action Eq. (\ref{action0}) can be written as%
\begin{equation}
I=\int dt\int d^{3}\vec{X}\ \mathcal{L}(  \frac{d\vec{r}}{dt},\nabla
_{X}\vec{r})  \label{Action_Lag}%
\end{equation}
with the Lagrangian density%
\begin{equation}
\mathcal{L}=-J\ \varepsilon(\frac{1}{\gamma J}n_{0}^{\ast
})\label{L_dens}%
\end{equation}
Note that this Lagrangian density does not depend explicitly on the field
$\vec{r}(\vec{X},t)$ but only on $d\vec{r}/dt$ and $\nabla_{X}\vec{r}$. The
symbol $d/dt$ means the time derivative in time for a fixed $\vec{X}$ and
$\nabla_{X}$ stands for the gradient with respect to $\vec{X}.$

To calculate the Euler-Lagrange equation,%
\begin{equation}
\frac{d}{dt}\frac{\partial\mathcal{L}}{\partial(d\vec{r}/dt )  }+\nabla_{X}(  \frac{\partial\mathcal{L}}{\partial(
\nabla_{X}\vec{r})  } )  =0 \label{Euler_Lagrange}%
\end{equation}
we use the following properties of the Jacobian $J.$%
\begin{equation}
\nabla_{X}\cdot\frac{\partial J}{\partial(  \nabla_{X}r^{i})
}=\nabla_{X}\cdot(  \nabla_{X}r^{j}\times\nabla_{X}r^{k})  =0,
\label{Jacob_1}%
\end{equation}
and%
\begin{equation}
\frac{\partial J}{\partial(  \nabla_{X}r^{i})  }\cdot\nabla
_{X}=J\frac{\partial}{\partial r^{i}}. \label{Jacob_2}%
\end{equation}
In the above, the components $(i,j,k)$ of the vector $\vec{r}$
should be taken to be cyclic. Thus,
\begin{eqnarray}
\nabla_{X}(  \frac{\partial\mathcal{L}}{\partial(  \nabla_{X}\vec
{r})  })   &=&\sum_{k}\nabla_{X}\cdot\left\{  n^{2}\frac
{\partial}{\partial n}\ \frac{1}{n}\varepsilon(  n)  \frac{\partial
J}{\partial(  \nabla_{X}\vec{r})  }\right\} \nonumber\\
&=&\frac{n_{0}}{n^{\ast}}\nabla_{\vec{r}}\ P, \label{Term_1}
\end{eqnarray}
and
\begin{eqnarray}
\frac{\partial\mathcal{L}}{\partial( d\vec{r}/dt )  }
&
=&-\frac{\partial\varepsilon}{\partial n}n^{\ast}\frac{\partial(
1/\gamma)  }{\partial( d\vec{r}/dt )  }\nonumber\\
&=&+\gamma\frac{d\vec{r}}{dt}\frac{\partial\varepsilon}{\partial n}n^{\ast
}\nonumber\\
&=& n_{0}\frac{\varepsilon+P}{n} \vec{u} \label{Term2}%
\end{eqnarray}
so that we get%
\begin{equation}
\frac{d}{dt}(  \frac{\varepsilon+P}{n}\ \vec{u} )  =-\frac
{1}{n^{\ast}}\nabla_{\vec{r}}\ P \label{Euler_Rel_2}%
\end{equation}
which is exactly the relativistic Euler equation, Eq.(\ref{EulerRel}). Here,
as before, we defined $P$ as
\[
P\equiv\frac{\partial\varepsilon}{\partial n}n-\varepsilon,
\]
given Eq.(\ref{e(n)}).

An important fact is that, the canonical momentum field of $\vec{r}(t,\vec
{X})$ is%
\begin{equation}
\vec{\pi}=\frac{\partial\mathcal{L}}{\partial(  d\vec{r}/dt)
}=\frac{\varepsilon+P}{n}\ \vec{u}, \label{canonicalp}%
\end{equation}
so that the Hamiltonian of the system is given by%
\begin{eqnarray}
H  &=&\int d^{3}\vec{X}\ n_{0}^{\ast}\left[  \frac{\varepsilon+P}{n}\ \vec
{u}\right]  \frac{d\vec{r}}{dt}-L\nonumber\\
&=&\int d^{3}\vec{x}\ \left[  (  \varepsilon+P)  \ \vec{u}%
^{2}+\varepsilon\right] \nonumber\\
&=&\int d^{3}\vec{x}\ T^{00}, \label{H_dense}%
\end{eqnarray}
\ which means that the Hamiltonian density in $x$ coordinates is given by the
$(  00)  $ component of the energy-momentum tensor, $T^{\mu\nu
}=(  \varepsilon+P)  u^{\mu}u^{\nu}-Pg^{\mu\nu}$, as expected.

Furthermore, the total momentum $\vec{P}$ of the system is given by%
\begin{eqnarray}
\vec{P}  &=&\int d\vec{X}\ n_{0}^{\ast}\left[  \frac{\varepsilon+P}{n}%
\ \vec{u}\right] \nonumber\\
&=&\int d^{3}\vec{x}\ (  \varepsilon+P)  \gamma\vec{u}%
\ \label{P_system}%
\end{eqnarray}
which is nothing but the integral of the space part of $T^{0i}=(
\varepsilon+P)  u^{0}u^{i}$. Thus, the expression from the definition of
the Hamiltonian density is:
\begin{equation}
\ T^{00}=(  \varepsilon+P)  \ \vec{u}^{2}+\varepsilon
\end{equation}
can be rewritten as%
\begin{equation}
T^{00}u_{0}+T^{0i}u_{i}-\varepsilon u^{0}=0.
\end{equation}
Note that this is exactly the 0-th component of the eigenvalue equation
Eq.(\ref{Eigen}).

\subsection{Coarse Graining and Hydrodynamic Modeling}

In the derivations above, we have assumed that the relation in Eq.
(\ref{e(n)}) (or the general case Eq. (\ref{e_geral})) is valid locally. This
is somewhat a very severe condition, depending on the definition of the
densities. For example, when we consider the hydrodynamical description of the
relativistic heavy ion reactions, differently from the normal hydrodynamic
applications for macroscopic fluids, the typical fluid element size can not be
taken too much smaller than the whole system and also the time scale of the
collective motion can not be much larger than that of the microscopic one. In
such case fluctuations and inhomogeneities from the microscopic level are
expected to be very large and relations like Eq. (\ref{e(n)}) becomes
difficult to be well-defined.

On the other hand, the derivation of relativistic hydrodynamics from the
variational procedure suggests that we may apply similar method for certain
averaged densities over some statistical ensemble of fluid elements and still
obtain the equations for average densities of conserved quantities. For
example, let us consider a set of (very large number of) particles, instead of
a truly continuum medium, which is also quickly moving. In this case, the
density $n_{0}^{\ast}$ is a sum of Dirac delta functions so that any smooth
relation with the energy density becomes meaningless. However, in most cases
we do not have very precise resolution neither in space nor in time. Then we
may introduce an averaged density distribution by introducing the smoothing
function $W_{h}(  \vec{x})  $ instead of $\delta$ function in
Eq.(\ref{dens}) and also take a certain time average, to define the smoothed
density distribution as
\begin{equation}
n^{\ast}(  \vec{x},t)  =\int dt^{\prime}\int d^{3}\vec{X}%
\ n_{0}^{\ast}(  \vec{X})  \ U_{\tau}(  t^{\prime}-t)
W_{h}(  \left\vert \vec{x}-\vec{r}(  t^{\prime};\vec{X})
\right\vert )  ,\label{nsmooth}%
\end{equation}
where $U_{\tau}(  t)  $ and $W_{h}(  x)  $ are smoothing
positive definite functions (in mathematics are also called mollifiers which are approximations to the identity), satisfying

\begin{eqnarray}
U_{\tau}(  t)   &  \rightarrow&0,\ \left\vert t\right\vert >\tau.\\
\int U_{\tau}(  t)  dt &=&1,\\
\lim_{\tau\rightarrow0}\int\varphi(  t)  U_{\tau}(  t)
dt &=&\varphi(  0)  ,
\end{eqnarray}%

\begin{eqnarray*}
W_{h}(  \vec{r})   &  \rightarrow&0,\ \left\vert \vec{r}\right\vert
>h.\\
\int W_{h}(  \vec{r})  d^{3}\vec{r} &=&1,\\
\lim_{h\rightarrow0}\int\varphi(  \vec{r})  W_{h}(  \vec
{r})  d^{3}\vec{r} &=&\varphi(  \vec{0})  ,
\end{eqnarray*}
where $\varphi$ is an arbitrary well-behaved function. A typical example of
$U$ and $W$ is the Gaussian distribution. The positive parameters $\tau$ and
$h$ are the scales of the time and space resolutions, respectively. We can
also define the 3-current,%
\[
\vec{j}^{\ast}(  \vec{x},t)  =\int dt^{\prime}\int d^{3}\vec
{X}\ n_{0}^{\ast}(  \vec{X})  \frac{d\vec{r}}{dt^{\prime}}(
t^{\prime},\vec{X})  U_{\tau}(  t^{\prime}-t)  W_{h}(
\left\vert \vec{x}-\vec{r}(  t^{\prime};\vec{X})  \right\vert
)  .
\]
We can see that these averaged density and current satisfy the continuity
equation,%
\begin{eqnarray*}
\frac{\partial n^{\ast}}{\partial t} &=&\int dt^{\prime}\int d^{3}\vec
{X}\ n_{0}^{\ast}(  \vec{X})  \ \frac{d}{dt}U_{\tau}(
t^{\prime}-t)  W_{h}(  \left\vert \vec{x}-\vec{r}(  t^{\prime
};\vec{X})  \right\vert )  \\
&=&-\int dt^{\prime}\int d^{3}\vec{X}\ n_{0}^{\ast}(  \vec{X})
\ \frac{d}{dt^{\prime}}U_{\tau}(  t^{\prime}-t)  W_{h}(
\left\vert \vec{x}-\vec{r}(  t^{\prime};\vec{X})  \right\vert
)  \\
&=&\int dt^{\prime}\int d^{3}\vec{X}\ n_{0}^{\ast}(  \vec{X})
\ U_{\tau}(  t^{\prime}-t)  \frac{d}{dt^{\prime}}W_{h}(
\left\vert \vec{x}-\vec{r}(  t^{\prime};\vec{X})  \right\vert
)  \\
&=&-\int dt^{\prime}\int d^{3}\vec{X}\ n_{0}^{\ast}(  \vec{X})
\ U_{\tau}(  t^{\prime}-t)  \frac{d\vec{r}}{dt^{\prime}}\cdot
\nabla_{x}W_{h}(  \left\vert \vec{x}-\vec{r}(  t^{\prime};\vec
{X})  \right\vert )  \\
&=&-\nabla_{x}\cdot\vec{j}^{\ast}\mathbf{.}%
\end{eqnarray*}

Using this conserved smoothed current and density, we can derive the
four-current,%
\[
j^{\mu}=(
\begin{array}
[c]{c}%
n^{\ast}\\
\vec{j}^{\ast}%
\end{array}
)
\]
and the proper density,%
\[
n=\sqrt{j^{\nu}j_{\nu}}.
\]
The smoothed four-velocity field is then defined as%
\[
u^{\mu}=\frac{1}{n}\ j^{\mu}.
\]
Once we defined the (smoothed) velocity field above, we can consider the local
rest frame at the space time point given by $x=(t,\vec{x})$ in terms of a Lorentz
transformation $\Lambda(  u)$, defined by
\[
\Lambda(  u)  u=(
\begin{array}
[c]{c}%
1\\
0\\
0\\
0
\end{array}
)  .
\]
We can also introduce the smoothed energy-momentum tensor in analogous way.
Let
\[
T_{M}^{\mu\nu}(\vec{r}(  R,t)  ,t)
\]
be the energy-momentum tensor associated to the matter at the space position
$\vec{x}=\vec{r}(  R,t)  $ at time $t.$ From this, we introduce the
smoothed energy-momentum tensor,%
\begin{equation}
T^{\mu\nu}(  \vec{x},t)  =\int dt^{\prime}\int d^{3}\vec{x}%
^{\prime}\ U_{\tau}(  t^{\prime}-t)  W_{h}(  \left\vert
\vec{x}-\vec{x}^{\prime}\right\vert )  T_{M}^{\mu\nu}(\vec{x}^{\prime
},t^{\prime}).\label{Tmunu_smooth}%
\end{equation}
As before, we can show that%
\[
\partial_{\mu}T^{\mu\nu}=\int dt^{\prime}\int d^{3}\vec{x}^{\prime}\ U_{\tau
}(  t^{\prime}-t)  W_{h}(  \left\vert \vec{x}-\vec{x}^{\prime
}\right\vert )  \partial_{\mu}^{\prime}T_{M}^{\mu\nu}(\vec{x}^{\prime
},t^{\prime})=0,
\]
if%
\[
\partial_{\mu}T_{M}^{\mu\nu}(x)=0.
\]
From this smoothed energy-momentum tensor $T^{\mu\nu},$we can calculate also
the smoothed proper energy-density as%
\begin{equation}
\varepsilon\equiv u_{\mu}u_{\nu}T^{\mu\nu}.\label{edens}%
\end{equation}

The smoothed proper energy-density defined in this way is an average of the
energy density viewed in the rest frame of the matter flow. The average is
taken over all contributions from the matter within the range of the
coarse-graining scale in space-time.

In terms of hydrodynamic modeling, we take these smoothed distributions as the
dynamical variables that represent the system. For one collisional event,
where every microscopic state is specified, we can calculate these
hydrodynamical variables. Inversely, however, for a given set of
hydrodynamical variables there may exist many different microscopic
configurations that give rise to the same set of hydrodynamical variables. Let
us call $\Omega$ the set of all collisional events with microscopic
configurations that produce a given profile of the four-current $j^{\mu}$ at
the initial time $t_{0}.$ In other words, $\Omega$ is the ensemble of events
that corresponds to an identical hydrodynamical profile. If we calculate
$\varepsilon$ at a given space-time point $x$ for each of the events in
$\Omega,$ they do not coincide in general. This is so even for $j^{\mu}(
\vec{x},t)  ,$ if $t>t_{0}.$ However, if the coarse-graining size is
large enough, the ensemble of microscopic matter which contribute to the
hydrodynamic variables at the point $x$ can be sufficiently large enough so
that we expect that $\varepsilon$ and $n$ will distribute sharply around their
respective mean-values, say $\bar{\varepsilon}$ and $\bar{n},$ due to the
central limit theorem. Furthermore, since $\bar{\varepsilon}$ and $\bar{n}$
are the average energy and matter densities belonging to the same fluid
element specified by the coarse-graining volume, we expect that they will have
a strong correlation, in such a way that they are functionally related
\begin{equation}
\bar{\varepsilon}=\bar{\varepsilon}(  \bar{n})  .\label{EoS}%
\end{equation}

When we neglect the fluctuations in $\varepsilon$ and $j^{\mu},$ we may think
of a physical model in terms of the averaged matter current
which defines $\bar{n}^{\ast}$ and velocity field $\mathbf{\bar{v}}$ as the
dynamical variable. For this, we may take the model action as%
\begin{equation}
I=-\int d^{4}x\ \bar{\varepsilon}(  \frac{1}{\gamma}\bar{n}^{\ast
})  .\label{ModelAction}%
\end{equation}
The variational principle for this action establishes the ideal hydrodynamic
equation of motion for the coarse-grained variables. This model describes the
time evolution of the system belonging to a given statistical ensemble
$\Omega,$ that is, the hydrodynamic initial condition.

When the fluctuations of hydrodynamical variables within the ensemble $\Omega$
are not negligible, the above ideal fluid modeling fails and we have to extend
the hydrodynamical variable as stochastic ones. Even in this case, it is
possible to construct the hydrodynamical model with the variational principle.
In Ref. \cite{1751-8121-45-25-255204}, it is shown that the variational principle for the
action%
\[
I=I(  \varepsilon,n^{\ast},\vec{v})
\]
extended to fluctuating (stochastic) variables leads to the Navier-Stokes
equation in the non-relativistic case. As expected, the fluctuation is
directly related to the dissipation coefficients.

Therefore  in the present vision, the viscosity is an unknown model parameter
related directly to the coarse-graining scale. It may take different values
than that obtained, for example, from lattice QCD calculations,
except for the limiting case of thermal equilibrium.
In fact, the simulations by ideal hydrodynamics with fluctuating
initial conditions give rise to a similar behavior of $p_t$ dependence of $v_2$
to that of the viscous hydrodynamics \cite{HAMA2005,PhysRevLett.101.112301}.

It is interesting to note that, although the coarse-graining procedure defined
through Eqs. (\ref{nsmooth}) and (\ref{Tmunu_smooth}) is neither local nor
covariant, the resulting equation from the variational principle is written as
if the theory is relativistically covariant\footnote{ This situation reminds
us the similar situation of the frequently used mean-field approximation of an
effective field theory. Even if the final form of the approximation keeps the
system of equations as if they are relativistic covariant, physically it is
not covariant. }.

\section{Event by Event Hydrodynamics}

As shown above, the ideal fluid description of heavy ion reaction can be
regarded as an effective model for the coarse-grained variables for the flow
based on the assumption, Eq.(\ref{EoS}) together with the model action,
Eq.(\ref{ModelAction}). In this vision, the success of ideal fluid modeling in
relativistic heavy ion collisions is, in short, reduced to the adequacy of the
approximations contained in Eqs. (\ref{EoS}) and (\ref{ModelAction}) for
collective variables.
These two equations allow
for much wider microscopic configurations than those restricted by
the ``local thermal equilibrium'' condition (as defined in the introduction),
usually considered as the necessary condition for the validity of a
hydrodynamic description.

As we lengthy described in the previous section, the definition of
hydrodynamic variables requires the coarse-graining procedure. The size of the
hidden ensemble $\Omega$ depends on the coarse-graining size. The larger the
size of coarse-graining is, the larger the ensemble $\Omega$ becomes. For
larger $\Omega,$ the two conditions, Eqs. (\ref{EoS}) and (\ref{ModelAction})
have larger chance to be satisfied and the probability of the success of ideal
fluid model increases. On the other hand, for larger coarse graining size, we
loose the resolution in the space-time recognition. In terms of hydrodynamic
variables, we will not be able to see the inhomogeneities in density profile that
has smaller wavelength than the coarse-graining scale. This affects directly
the class of observables that the model can describe. The extreme example is
when the coarse-graining size is larger than the system size and time
evolution. Then the ensemble $\Omega$ can be regarded as the statistical
ensemble of the whole system, and the resultant system is a simple fireball
model. Therefore, the thermal model for the particle ratio can be considered
in this category.

The coarse-graining size is thus intimately related to the class of
observables and the validity of hydrodynamic description. For some observables
which do not require a precise space-time resolution of the system,
hydrodynamics description with a large coarse-graining scale maybe sufficient.
In such cases, the real local thermal equilibrium is not necessary for the
success of the hydrodynamic description. In addition, the experimental
observables are usually averaged over collision events classified in terms of
configurations rather loosely defined, such as centrality, event
plane, etc. This may mask the important information contained in each
hydrodynamic evolution. For example, the recent studies \cite{PTPS.193.315,GDenicol}
of hydrodynamics with fluctuating initial condition on the elliptic flow
$v_{2}$ indicate that the linear relation of integrated $v_{2}$ to the initial
deformation parameter $\epsilon_{2}$ are quite insensitive not only to the
granularity in the initial condition but also to the transport coefficients,
when averaged over many events. This means that the event averaged integrated
$v_{2}$ has a poor resolution for the hydrodynamic evolutions, so that the
coarse-graining size of the respective hydrodynamics needs not to be small.

In order to claim that the real \emph{local} hydrodynamics is valid, we need
to have observables that reflect the genuine hydrodynamic profile in event by
event basis. For example, remnant of a sharp shock wave propagation, if exist,
would be a good observable which tells the possible coarse-graining size of
the collective flow, since a shock wave \cite{Stoecker2005121,CasalderreySolana2006577,PhysRevC.78.034901} is a genuine local
hydrodynamic phenomena. The shock thickness should not be larger than the
coarse graining scale of the collective flow. Another example of genuine
hydrodynamic signal may be the emergence of Kelvin-Helmholtz instability in
longitudinal direction as suggested in \cite{PhysRevC.85.054901}.

In any case, it is extremely important to study the hydrodynamic observables
in event-by-event basis, as proposed a long time ago \cite{PhysRevC.55.1455,Aguiar2002639,PhysRevLett.97.202302,ANDRADE2007}. Recently,
thanks to the high quality experimental data, information contained in
event-by-event fluctuations in hydrodynamic analysis and their physical origin are being more actively
investigated \cite{PhysRevLett.103.242301,Wang:2012fx,Qiu2012151,PhysRevC.81.054905,PhysRevC.83.064904,Schenke:2012wb,PhysRevC.85.034907}.

The key point is that when we apply the hydrodynamic modeling, we do not know
a-priori the coarse-graining scale suitable for the flow variables in the real
scenario\footnote{Usually we estimate this in terms of Knudsen number or
mean-free path, but they are useful when the system is composed of gas.}.
Formally we can apply the hydrodynamic description even to very inhomogeneous
initial conditions. But this requires a correspondingly small coarse-graining
scale so that the hydrodynamic description breaks down eventually. Thus, the
question is to know how far we can infer the initial condition inhomogeneity
within the hydrodynamic modeling without knowing the coarse-graining scale
actually permitted in the process.

In order to clarify the question of the validity of hydrodynamic models and
the coarse-graining scale, it is essential to find out the set of observables
which carry the information of the dynamical effects coming from the initial
inhomogeneities. For this purpose, we take the following strategy. First,
prepare initial conditions with different degree of inhomogeneities in a
controlled way and study the flow dynamics from these initial conditions within
an ideal fluid modeling. That is, we assume that the hydrodynamic coarse
graining scale is smaller than the inhomogeneity of the given initial
condition. Then, look for the final state observables which reflect directly
the scale of inhomogeneity in the initial condition. Once such observables can
be identified, we may look for these in experimental data. If experimental
data show such observables which characterize the scale of initial state
granularities, we may conclude that the coarse-graining size which still
permits the fluid-dynamical description should be of the order of
corresponding granularity scale revealed in the data. Furthermore, observation
of such granularity observables indicate strongly the granularities in the
initial condition, which should be useful clues for the initial state of
dynamics created by QCD.

\subsection{Effects of granularity in the initial conditions}

As previously discussed, we are not able to know {\it a priori} the coarse-graining scale required for the hydrodynamic description of
heavy ion collisions.
On the other hand, the scale is intimately related to the space-time resolution.
Therefore, if the experimental data reveals a signal that indicate the granularity scale of the initial condition in terms of the hydrodynamic variables, we can infer that the corresponding resolution, and consequently the coarse-graining scale is compatible to this resolution.
In this sense, it is very important to establish such a measure which reflects the granularity scale in the initial condition within the framework of hydrodynamic modeling.
For this purpose we investigate how the granularity of the initial condition affects the flow observables.

In order to analyze the granularity we propose a para\-metrization of the initial conditions based on the Glasma picture.
Each event is described as a set of overlapping longitudinal tubes where the total energy content, $E_t$, and their width, $\sigma$, are taken as parameters \cite{Mota2011188,PTPS.193.315}.
The tubes are distributed according to a wounded nucleons distribution with the constants chosen to describe Au-Au collisions at $\sqrt{s_{\rm NN}}=200$GeV/$A$.

In table~\ref{tab:params} we show all the values that were used during our simulations.
For each parameter set, we computed the 2D boost invariant hydrodynamic evolution of 2000 events.
The impact parameter range was chosen as $b=$0--12fm and centrality classes were computed according to the final thermal multiplicity of pions, $N_\pi$.
For the sake of simplicity, we compare the results of the extreme cases denoted A, B, C, and D (with decreasing level of granularity).
The other simulations are used later when we systematically analyze the parameter dependency.

The granularity of the initial conditions generated by this model is controlled by the two parameters that give the shape of the transverse Gaussian distribution of each tube: $\sigma$ and $E_t$.
More conveniently, one may also think of the peak energy density of the tube, $e=E_t/(2\pi \sigma^2)$.
Higher granularities are obtained by higher $e$ and smaller $\sigma$.

\begin{table}
\begin{center}
\caption{Sets of parameters used for our simulations. The quantities in the header are: the tube width, $\sigma$, the total transverse energy content, $E_t$, the number of tubes for a central collision ($b=0$), $N^{\rm max}$, and the peak energy density of the tubes, $e$.}
\label{tab:params}
\begin{tabular}{c|c|c|c|c}
& $\sigma$[fm] & $E_t$[GeV/fm] & $N^{\rm max}$ & $e$[GeV/fm$^3$] \\
\hline
D & 1 & 2.6 & 1000 & 0.41 \\
& 1 & 6.5 & 400 & 1.0 \\
C & 1 & 13 & 200 & 2.1 \\
& 0.7 & 2.6 & 1000 & 0.84 \\
& 0.7 & 6.5 & 400 & 2.1 \\
& 0.7 & 13 & 200 & 4.2 \\
& 0.5 & 2.6 & 1000 & 1.7 \\
& 0.5 & 6.5 & 400 & 4.1 \\
& 0.5 & 13 & 200 & 8.3 \\
B & 0.35 & 2.6 & 1000 & 3.4 \\
& 0.35 & 6.5 & 400 & 8.4 \\
A & 0.35 & 13 & 200 & 17 \\
\end{tabular}
\end{center}
\end{table}

\begin{figure}
\begin{center}
\includegraphics{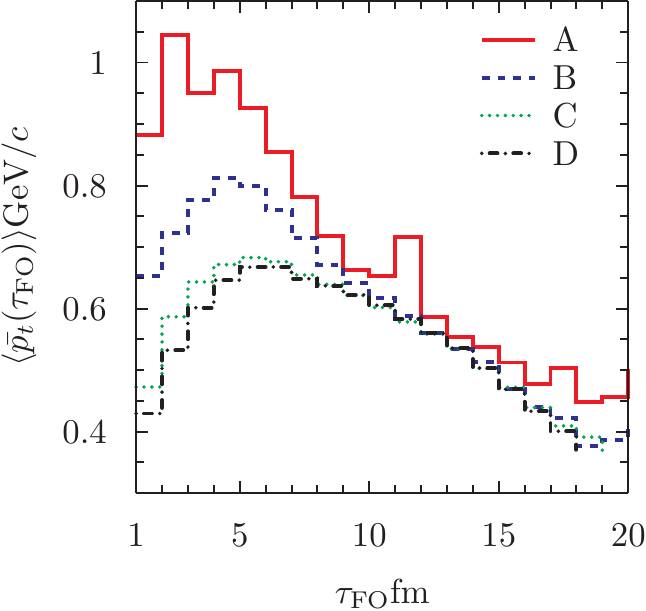}
\caption{Mean transverse momentum dependency on the emission time for different granularities.}
\label{fig:meanptau}
\end{center}
\end{figure}

In these calculations, we observe that there is a correlation between the averaged mean-$p_t$ of final particles, $\ave{\bar{p}_t}$, and the instant of their freezeout, $\tau_{\rm FO}$.
Figure~\ref{fig:meanptau} shows that the high-$p_t$ particles tend to be emitted earlier than those of low $p_t$, except for the very early times.
Furthermore, this feature is enhanced by the granularity.

The result in fig.\ref{fig:meanptau} can be physically understood as follows.
The rise of $\ave{\bar{p}_t}$ at early times is due to the initial acceleration of the fluid whose transverse velocity is set to zero at $\tau_0=1$fm.
On the other hand, the decrease of $\ave{\bar{p}_t}$ in later times indicates that the bulk of fluid is subjected to smaller pressure gradients than the fluid elements present in the periphery -- which freezeout earlier.
Therefore, within the hydrodynamic picture, one may expect that the information about the space-time evolution of the system can be
extracted from the $p_t$ dependence of flow observables.

\subsection{\nexspherio}

Following the idea in the previous section, we propose to calculate the flow observables for separate $p_t$ regions.
We chose two $p_t$ regions denoted as low for $p_t=$0--0.5GeV/$c$ and high for 0.5--5 GeV/$c$.
In this section, we present calculations using the \nexspherio  model \cite{PhysRevC.65.054902,Aguiar2002639,HAMA2005}.

As argued before, we expect the initial geometry to be more strongly reflected in early emitted particles -- which tend to have higher $p_t$ -- than in late emitted particles -- lower $p_t$.
To check this hypothesis in a full 3D model, we have computed the flow and eccentricity phases, respectively,
$$
\psi_n^{`A} = \frac{1}{n} `arctan \frac{ \sum_{i; p_t\in `A} `sin(n\phi)  }{ \sum_{i; p_t\in `A} `cos(n\phi)  },
$$
and
$$
\Psi_n = \frac{1}{n} `arctan \frac{ \int rdrd\phi\, e(r,\phi) r^2 `sin(n\phi)  }{ \int rdrd\phi\, e(r,\phi) r^2 `cos(n\phi)  },
$$
where $n$ is the harmonic number, A is the $p_t$ range and $e(r,\phi)$ is the energy density distribution of the fluid in the initial condition.
We are interested in the relative phases between the low- and high-$p_t$ and $\Psi_n$.
For the sake of brevity, we will denote $\Delta\psi_n^{`A} = \abs{ \psi_n^{`A} - \Psi_n }$.

\begin{figure}
\begin{center}
\includegraphics[height=.23\textheight]{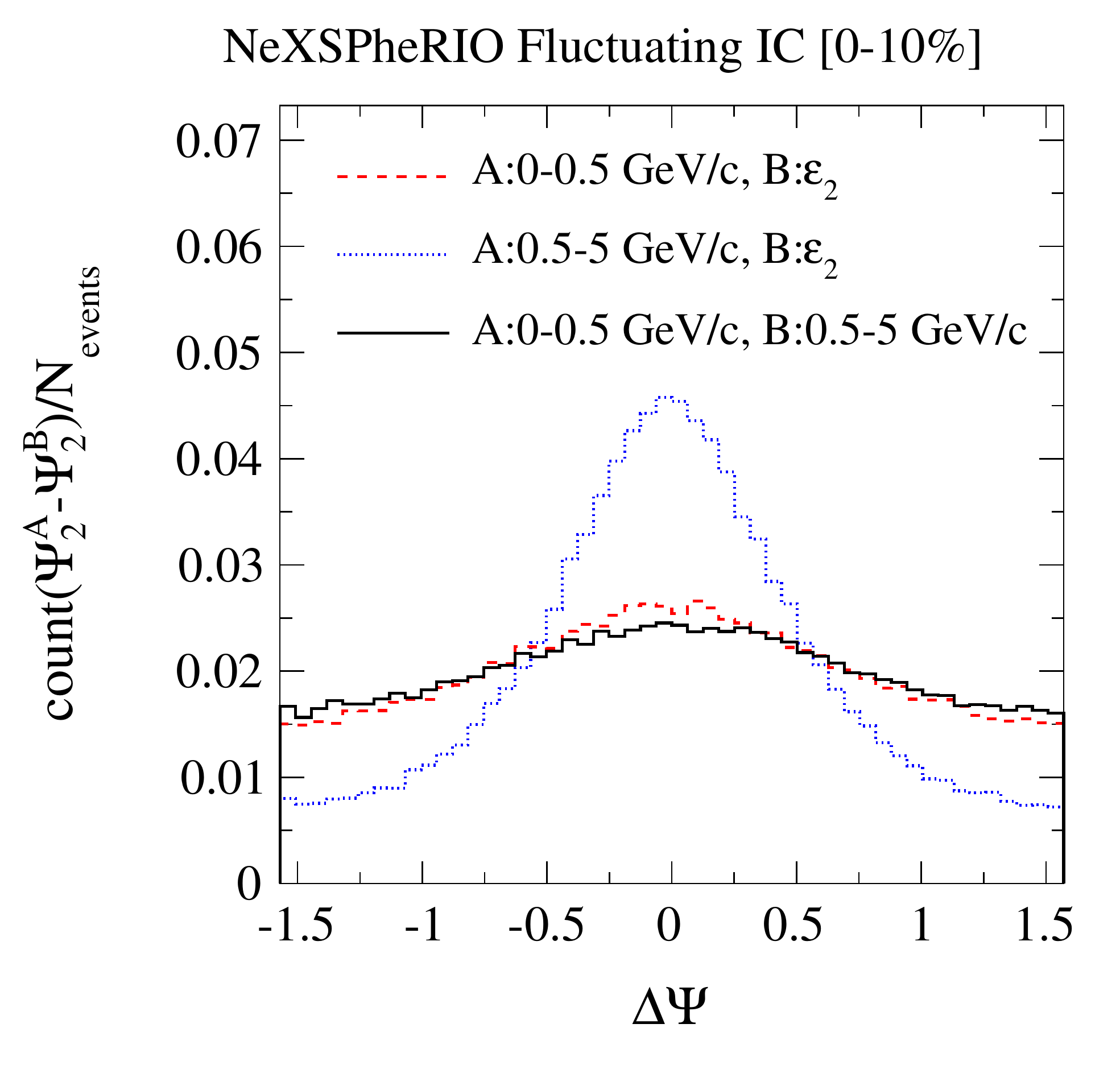}
\includegraphics[height=.23\textheight]{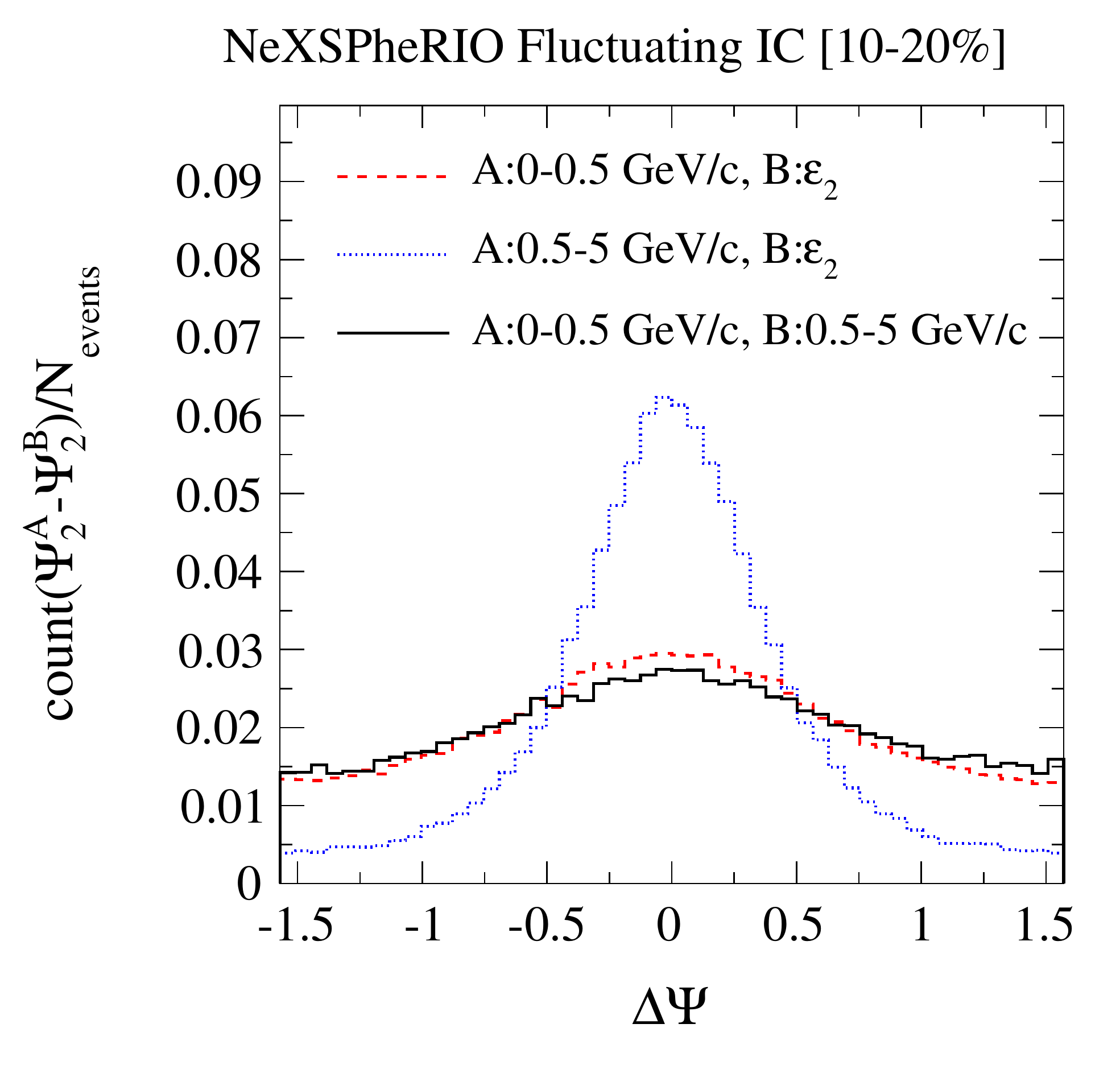}
\includegraphics[height=.23\textheight]{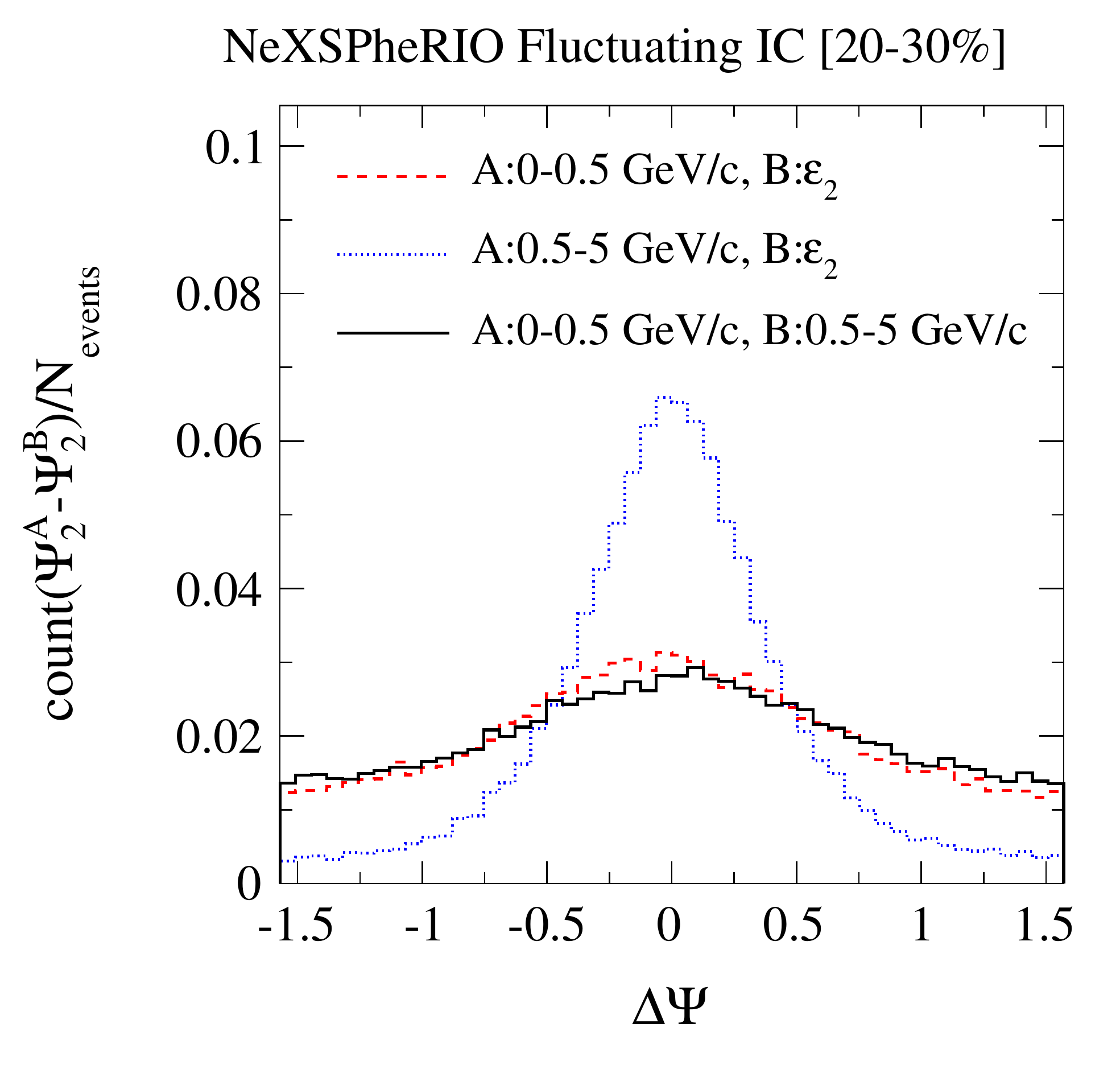}
\includegraphics[height=.23\textheight]{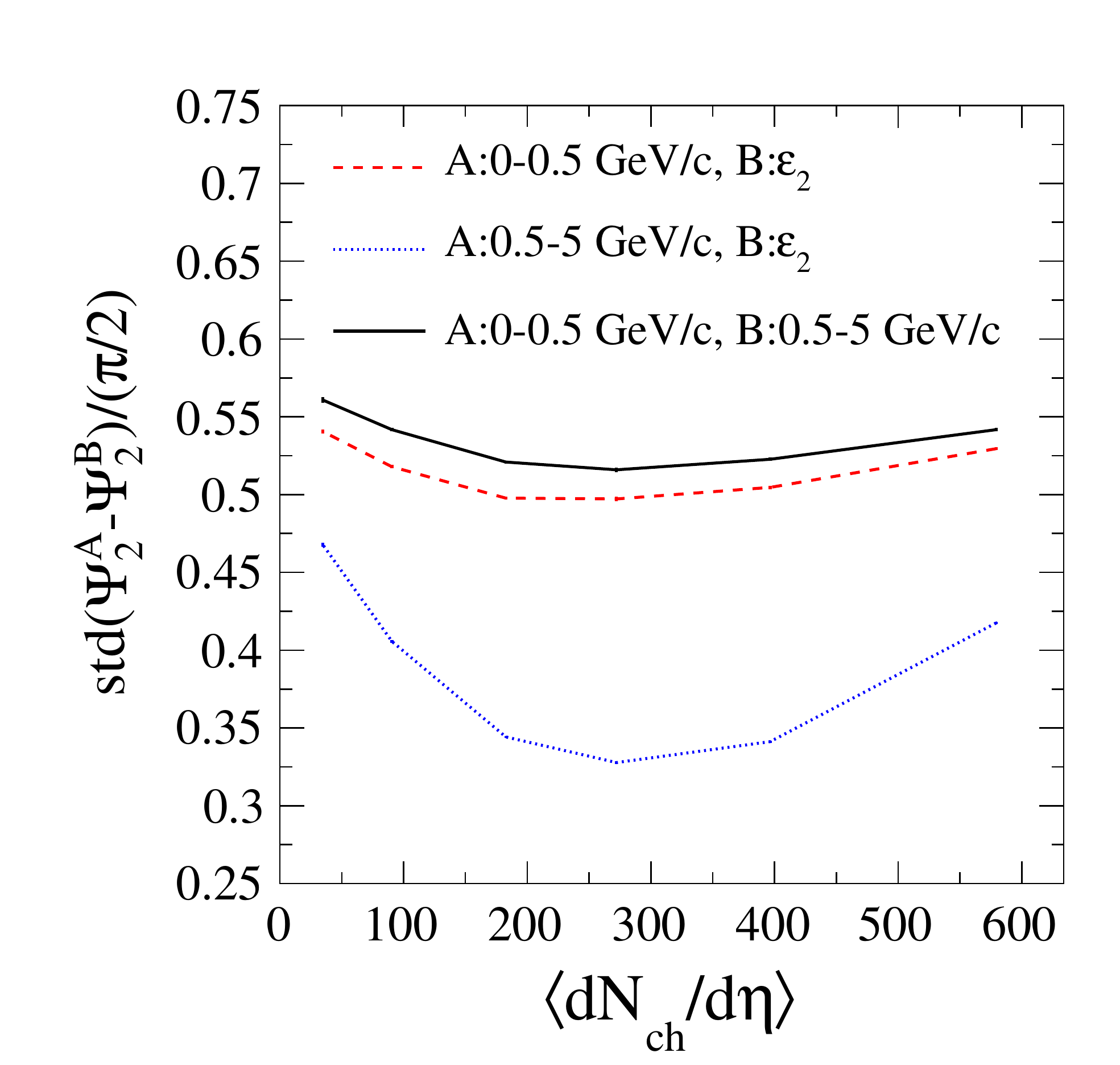}
\caption{NexPSheRio results for the relative phase distributions (explained in the text).}
\label{fig:spherio}
\end{center}
\end{figure}

In fig.\ref{fig:spherio} we show the distributions of $\Delta\psi_2^{\rm high}$ (dashed) and $\Delta\psi_2^{\rm low}$ (dotted).
The top three plots show the distribution for 0--10\%, 10--20\% and 20--30\% centrality bins, respectively.
The distribution of $\Delta\psi_2^{\rm low}$ is peaked at $\Delta\psi=0$ but is very broad and spans over all the possible values $0$--$\pi/2$.
On the other hand, $\Delta\psi_2^{\rm high}$ is much more concentrated around $\Delta\psi=0$.
These results confirm our expectation base on the result of fig.\ref{fig:meanptau}.
We observe that the high-$p_t$ particles reflect better the properties of the initial condition when compared to the low-$p_t$ particles.

The centrality dependency of the standard deviation of the distributions, $\std{\Delta\psi_2^{\rm high(low)}}$, are shown in the bottom plot in fig.\ref{fig:spherio}.
Note that $\std{\Delta\psi_2^{\rm high}}$ (normalized by the maximum value of $\pi/2$) reaches a minimum of 0.33 for semi-central collisions (20--30\%) and is consistently lower than $\std{\Delta\psi_2^{\rm low}}$.

Now we use the fact that $\psi_2^{\rm high}$ is closely related to $\Psi_2$ in order to experimentally estimate $\abs{ \psi_2^{\rm low} - \Psi_2 }$ by substituting $\Psi_2$ by the experimentally measurable quantity $\psi_2^{\rm high}$.
The solid lines in the plots if fig.\ref{fig:spherio} represent the distribution $\Delta\psi_2^{\rm high|low} = \abs{ \psi_2^{\rm high} - \psi_2^{\rm low} }$.
Note that this quantity is in turn an experimental observable.

As expected, the distribution of $\Delta\psi_2^{\rm high|low}$ (solid) follows closely that of $\Delta\psi_2^{\rm low}$ (dotted).
This indicates that one can extract experimentally the difference between the phase of the late emitted particles and the initial phase -- by means of the early emitted particles.
Furthermore, this difference depends on the details of space-time evolution,
in particular depends on the level of granularity of the initial condition.

\subsection{Granularity dependence}

In this section we perform a systematic study of the dependence of $\Delta\psi_2^{\rm high|low}$ with respect to the granularity, by changing the model parameters $\sigma$ and $E_t$.

The distributions for the four parameter sets (labeled A, B, C and D in table\ref{tab:params}, with decreasing level of granularity) are presented in fig.\ref{fig:n2}.
For the most smooth case, D, there is no phase shift during the space-time evolution and almost all events fall in the central bin, $\Delta\psi=0$.
On the other hand, for the case A, where the initial condition is the most granular among the cases we studied, there is a broad distribution of $\Delta\psi_2^{\rm high|low}$.
This result confirms that this observable is indeed very sensitive to the initial condition granularity.
Furthermore, as in the \nexspherio\ scenario, $\Delta\psi_2^{\rm high|low}$ follows $\Delta\psi_2^{\rm low}$.

\myfigure{
\includegraphics[height=.23\textheight]{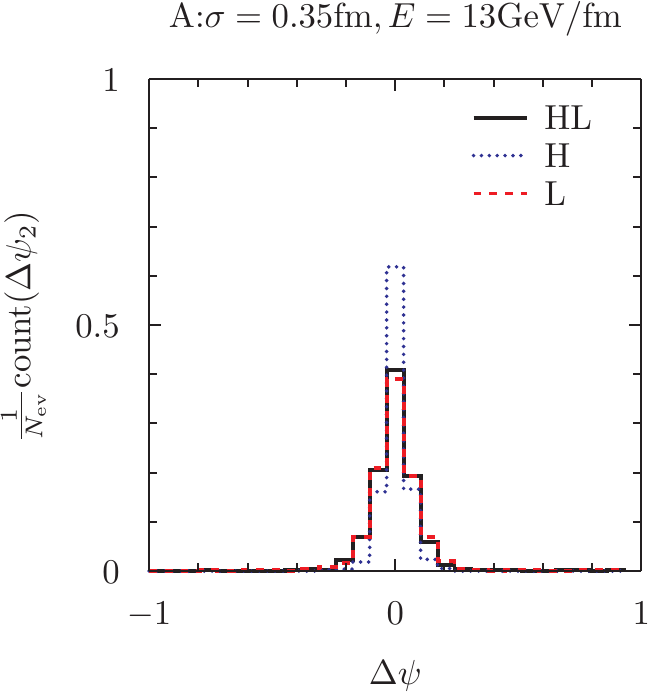}
\includegraphics[height=.23\textheight]{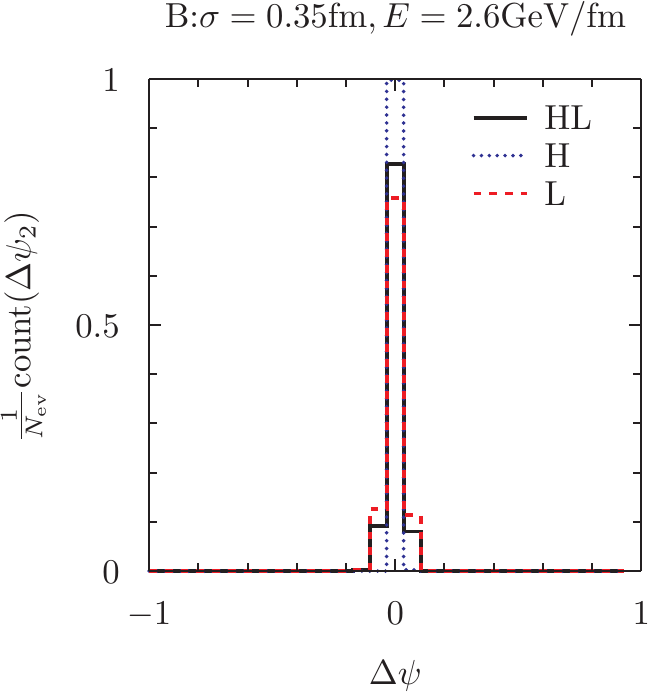}
\includegraphics[height=.23\textheight]{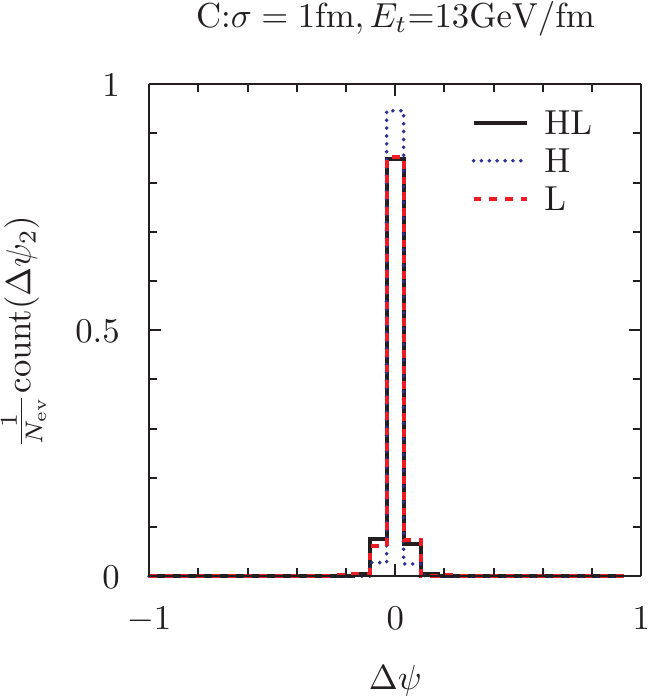}
\includegraphics[height=.23\textheight]{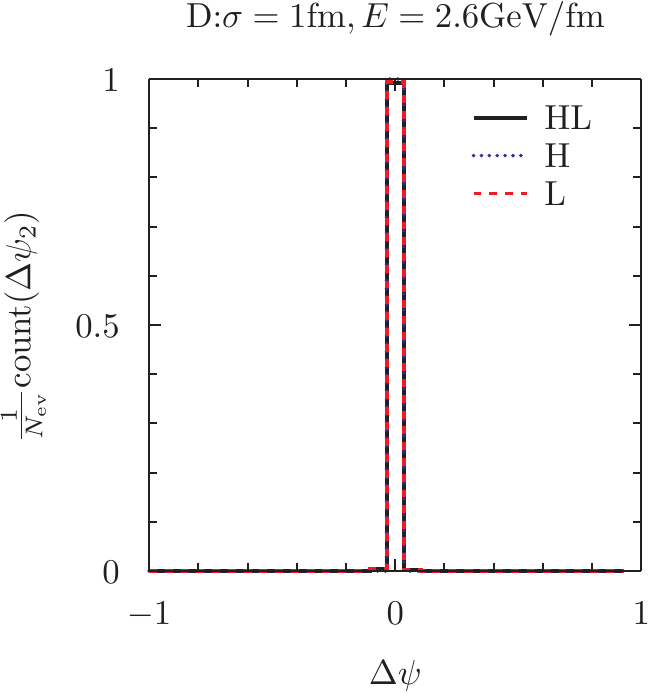}
\caption{
Relative phase distributions for $\Delta\psi_2^{\rm high|low}$ (HL), $\Delta\psi_2^{\rm high}$ (H),$\Delta\psi_2^{\rm low}$ (L).}
\label{fig:n2}
}

We also compute $\std{\Delta\psi_2^{\rm high|low}}$ for all values in table \ref{tab:params}.
The result is shown in the first plot of fig.\ref{fig:sigmadep} as a function of the tube width, $\sigma$, for different $E_t$.
There is a clear granularity dependence of the observable.
Higher granularities -- small $\sigma$ and large $E_t$ -- produce broader distributions of $\std{\Delta\psi_2^{\rm high|low}}$.
This suggests that the experimental analysis of this observable may provide important information about the geometrical properties of the initial condition.

It is also argued that higher harmonics would be sensitive to smaller angular variation, thus, within this approach, we could expect a higher sensitivity to the granularity.
However, since the observable proposed here is also affected by the evolution dynamics and the different freeze-out times, the result might not show a clear dependence.
The granularity dependence for the higher harmonics is shown in fig.\ref{fig:sigmadep}.
The higher harmonics have larger width than the elliptic case, showing that they are more sensitive for the space-time resolution, as expected. We also observe that, there is a difference in the dependence to the variation of tube energy content $E_t$.
In the case of odd harmonics, there is no difference between the results of different $E_t$ configurations, whereas, in the case of even harmonics, the curves are separate.
This distinction between the behavior observed for odd and even harmonics is still under investigation.

%

\myfigure{
\includegraphics[height=.23\textheight]{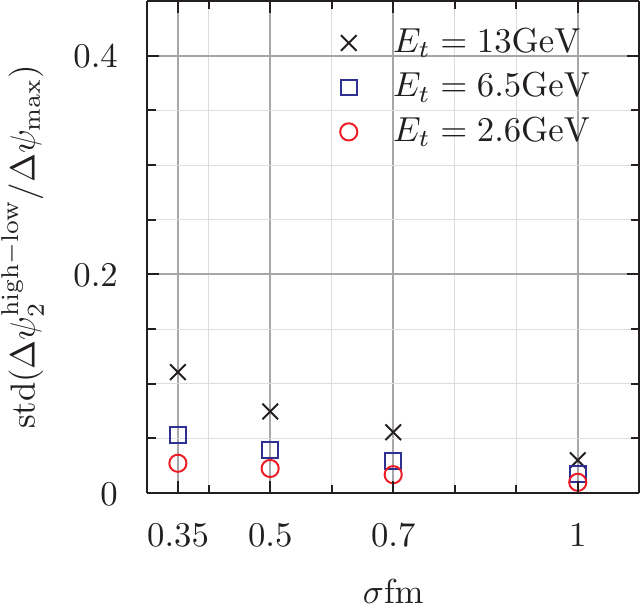}
\includegraphics[height=.23\textheight]{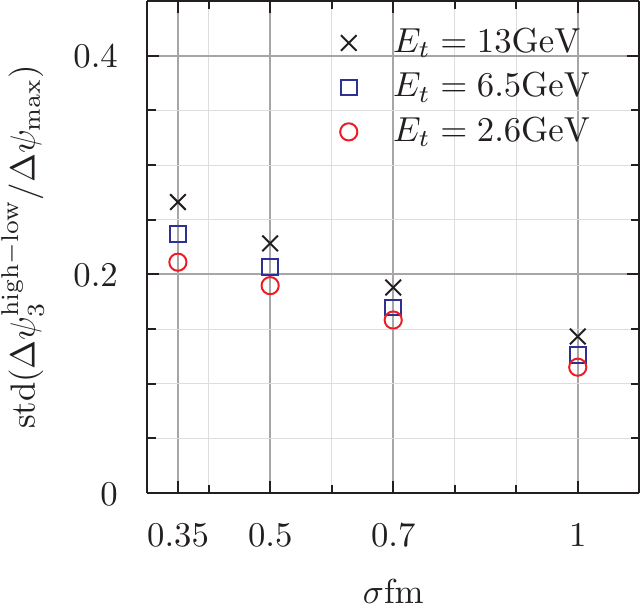}
\includegraphics[height=.23\textheight]{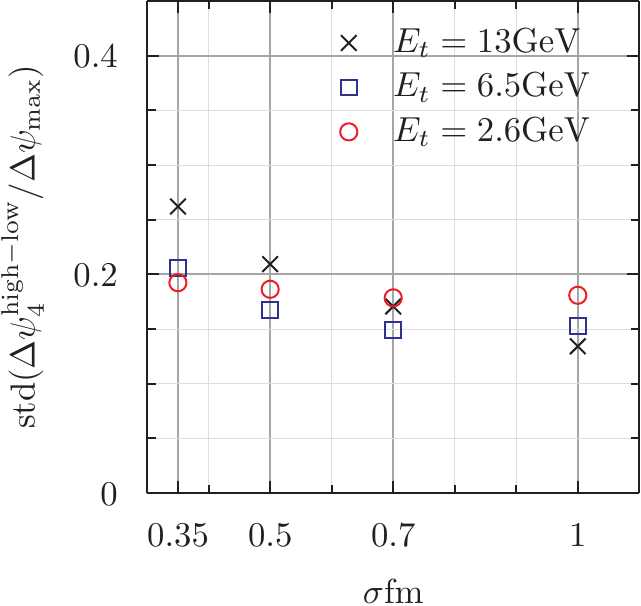}
\includegraphics[height=.23\textheight]{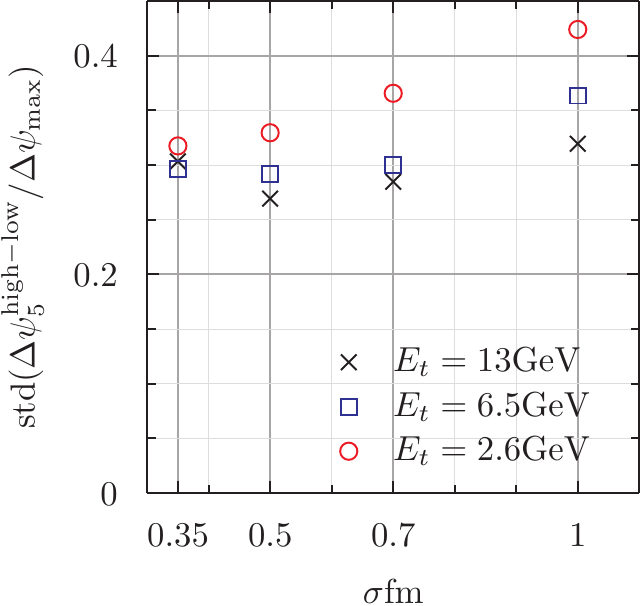}
\caption{Granularity dependence of $\std{\Delta\psi_2^{\rm high|low}}$.}
\label{fig:sigmadep}
}

\section{Discussion and Outlook}

In this paper, we discussed the meaning of hydrodynamic modeling for the
process of relativistic heavy ion collisions. We argued that it can be
considered as an effective model defined by Eq.(\ref{ModelAction}) for coarse
grained hydrodynamic variables. There, the validity of Eq.(\ref{EoS}) is a key
point when the average over the ensemble $\Omega$ is taken.

As mentioned, the most important issue is that for the hydrodynamic modeling,
we don't know a-priori the valid coarse graining scale. The final form of
relativistic hydrodynamics after the coarse-graining is a set of equations of
local fields and  does not have any information of the coarse-graining scale.
Even if the coarse-graining size for the validity of Eq. (\ref{EoS}) should be
large, we may model the system as if the coarse-graining size is very small.
Furthermore, depending on the observable, they can be very insensitive to the
fluctuations in each event to extract the detailed dynamical
information \cite{GDenicol}. Some observables carry only information which are
less affected by the coarse-graining size. On the contrary, if the behavior of
an observable is well described by the hydrodynamic modeling, we can conclude
that at least the coarse-graining size should be compatible with the
observable. In this sense, the success of ideal hydrodynamics for the
description of elliptic flow parameter $v_{2}$ tells us that the hydrodynamic
coarse-graining scale must be less than that of specified by this observable.

Some specific combinations of observables analyzed in event-by-event basis can carry more
detailed information on the hydrodynamic evolution of the system and many
efforts are now being done using higher harmonics \cite{PhysRevC.85.024901,PhysRevLett.103.242301}.
In the previous section, we showed
another example. There, if the hydrodynamic scenario is valid even for a
very spiky initial condition, particles with high momentum can be associated
to the early stage of the evolution. These particles are found to be emitted
from hot spots close to the surface of the system, since the time scale of
expansion of such hot spot is very short. Therefore, we may consider the
momentum of observed particles as a measure of time in hydrodynamic evolution.
In fact, the correlations between event planes defined in momentum bins
separated by the average $p_{t}$ seems very sensitive to the inhomogeneity
scale of the initial condition, although, unfortunately the present analysis
does not indicate still how to determine the inhomogeneity measure of the
initial condition from the experimental data. However, if these observables in
real experimental data, together with all other collective flow observables
are consistent to a hydrodynamic model, we could say that the coarse-graining
size is at least smaller than the inhomogeneity measure determined in this
way.

The authors would like to thank a fruitful discussion and suggestions of
G.Denicol. T.K. wishes to express his thanks for the hospitality of Prof. H.
Stöcker and Prof. D. Rischke. This work has been supported by CNPq, CAPES, FAPERJ and FAPESP from Brazil and
Theoretical Physics Department of the University of Frankfurt.

\bibliographystyle{unsrt}
\bibliography{references}

\end{document}